\documentclass[final,leqno,onefignum,onetabnum]{article}

\usepackage{amsmath}
\usepackage{amssymb}
\usepackage{hyperref}
\usepackage{graphicx}
\usepackage[usenames,dvipsnames]{xcolor}
\usepackage{amsfonts}
\usepackage{marginnote}
\usepackage{color}
\usepackage{bm}
\usepackage{cancel}
\usepackage{url}
\usepackage{oldgerm}
\usepackage{wrapfig}
\usepackage{lscape}
\usepackage{rotating}
\usepackage{wasysym}
\usepackage{soul}
\usepackage{verbatim}
\usepackage{mathrsfs}
\usepackage{booktabs}
\usepackage{geometry}

\newcommand{\mcB}{\mathcal{B}}

\newcommand{\mcF}{\mathcal{F}}
\newcommand{\mcI}{\mathcal{I}}

\newcommand{\mcL}{\mathcal{L}}

\newcommand{\mcY}{\mathcal{Y}}

\newcommand{\msH}{\mathscr{H}}
\newcommand{\mbR}{\mathbb{R}}
\newcommand{\mbRd}{{\mathbb{R}^d}}

\newcommand{\mbP}{\mathbb{P}}
\newcommand{\mbN}{\mathbb{N}}

\newcommand{\omg}{{\Omega}}

\newcommand{\beq}{\begin{equation}}
\newcommand{\eeq}{\end{equation}}

\def \ib{\mathbf{i}}
\def \lb{\mathbf{l}}

\def \xb{{\bf x}}
\def \yb{\bm{y}}

\def \Ub{\mathbf{U}}
\def \Fb{\mathbf{F}}
\newcommand*\samethanks[1][\value{footnote}]{\footnotemark[#1]}

\newtheorem{remark}{Remark}[section]

\title{Surrogate-based Ensemble Grouping Strategies for Embedded Sampling-based Uncertainty Quantification\footnote{Sandia National Laboratories is a multimission laboratory managed and operated by National Technology and Engineering Solutions of Sandia, LLC., a wholly owned subsidiary of Honeywell International, Inc., for the U.S. Department of Energy's National Nuclear Security Administration under contract DE-NA-0003525.}} 

\author{M. D'Elia\thanks{Center for Computing Research, Sandia National Laboratories, Albuquerque, NM ({mdelia, etphipp, msebeid@sandia.gov})} \and E. Phipps\samethanks \and A. Rushdi\thanks{University of California, Davis, CA ({aarushdi@ucdavis.edu})} \and M. S. Ebeida\samethanks[2]}

\begin{document}
\maketitle

\begin{abstract}
The embedded ensemble propagation approach introduced in~\cite{Phipps_2015} has been demonstrated to be a powerful means of reducing the computational cost of sampling-based uncertainty quantification methods, particularly on emerging computational architectures.  A substantial challenge with this method however is ensemble-divergence, whereby different samples within an ensemble choose different code paths.  This can reduce the effectiveness of the method and increase computational cost.  Therefore grouping samples together to minimize this divergence is paramount in making the method effective for challenging computational simulations.  In this work, a new grouping approach based on a surrogate for computational cost built up during the uncertainty propagation is developed and applied to model diffusion problems where computational cost is driven by the number of (preconditioned) linear solver iterations.  The approach is developed within the context of locally adaptive stochastic collocation methods, where a surrogate for the number of linear solver iterations, generated from previous levels of the adaptive grid generation, is used to predict iterations for subsequent samples, and group them based on similar numbers of iterations.  The effectiveness of the method is demonstrated by applying it to highly anisotropic diffusion problems with a wide variation in solver iterations from sample to sample.  It extends the parameter-based grouping approach developed in~\cite{DElia:2016} to more general problems without requiring detailed knowledge of how the uncertain parameters affect the simulation's cost, and is also less intrusive to the simulation code.
\end{abstract}


\pagestyle{myheadings}
\thispagestyle{plain}
\markboth{SURROGATE-BASED GROUPING STRATEGIES}{SURROGATE-BASED GROUPING STRATEGIES}

\section{Introduction}\label{sec:intro}
During the last decade the quantification of the uncertainty in predictive simulations has acquired great importance and is a topic of very active research in large scale scientific computing; we mention, e.g., random sampling methods~\cite{Fishman_96,Helton_Davis_03,McKay_79,Metropolis_49,Niederreiter:1978wr}, stochastic collocation~\cite{Babuska:2007kv,Nobile:2008dr,Nobile:2008kv,Xiu:2005ia} and stochastic Galerkin methods~\cite{Ghanem:1990p7167,Ghanem_Spanos_91,Xiu:2002p919}. It is very often the case that the source of uncertainty resides in the parameters of the mathematical models that describe phenomena of interest; when these parameters belong to a high-dimensional space or when the solution exhibits a non-smooth or localized behavior with respect to those parameters, sampling methods for uncertainty quantification (such as those mentioned above) may require a huge number of samples making the problem computationally intractable. For this reason, many methods, with the goal of reducing the number of samples, have been developed; among the others, we have locally adaptive sampling methods~\cite{Galindo_15,Gunzburger_14b,Stoyanov_15a}, multilevel methods~\cite{Barth_12,Barth_13,Barth_11,Cliffe_11,Giles_08}, compressed sensing~\cite{Doostan_11,Mathelin_12}, and tensor methods~\cite{Babuska:2007kv,Babuska_04,Back_11,Cohen_11,Frauenfelder_05,Ganapath_07,Nobile:2008dr,Nobile:2008kv,Roman_06,Xiu:2005ia}.

Nevertheless, the problem remains that for large scale scientific computing most of the computational cost is in the sample evaluation which, in most cases, corresponds to the numerical solution of a partial differential equation (PDE). Previous work~\cite{Phipps_2015} demonstrated that solving for groups (ensembles) of samples at the same time through forward simulations can dramatically reduce the cost of sampling-based uncertainty quantification (UQ) methods. However, fundamental to the success of this approach is the grouping of samples into ensembles to further reduce the computational work; in fact, the total number of iterations of the ensemble system is usually strongly affected by which samples are grouped together. 

In a previous work~\cite{DElia:2016} we investigated sample grouping strategies for local adaptive stochastic collocation methods applied to highly anisotropic diffusion problems where the uncertain diffusion coefficient is modeled by a truncated Karhunen-Lo\'eve (KL) expansion. There, we investigated PDE-dependent and location-dependent grouping techniques and demonstrated that a measure of the total anisotropy of the diffusion coefficient provides an effective metric for grouping samples as it is a good proxy for the number of iterations associated with each sample. We referred to this approach as parameter-based as it depends on the parameters, or coefficients (the diffusion tensor in this case), of the PDE. However, accessing problem-related information can be non-trivial or time/memory consuming. 

The main contribution of this follow-up work is the design of new grouping strategies that are cheaper and independent of the PDE. In the context of adaptive selection of the samples in the parameter space we propose a grouping strategy based on the construction of a (polynomial) surrogate for the number of solver iterations; the key idea of this approach is to group together samples with a similar predicted number of iterations. At each level of the adaptive grid generation algorithm we utilize data at previous levels to build surrogates of increasing accuracy so to maximize the computational saving. For the construction of the surrogate we consider standard polynomial sparse grid surrogates (SGS). This technique proves to be as successful as the parameter-based strategy while being less intrusive and requiring less computational work.

The use of surrogates is certainly not new in UQ methods for the solution of a variety of high-dimensional problems; we mention computational mechanics~\cite{breitkopf2013multidisciplinary,hao2012surrogate,rikards2004surrogate}, computational fluid dynamics~\cite{forrester2006optimization,razavi2012review}, microwave circuit design~\cite{bakr2000space,bandler2003based,couckuyt2010surrogate}, as well as system reliability and failure analysis~\cite{bichon2011efficient,POFDarts,li2011efficient,li2010evaluation}. 
Surrogate models have received more research attention as cheap approaches to deal with model uncertainties in scientific computing applications. More specifically, surrogate models (also known as emulators, metamodels, or response surface models) would fit a set of expensive evaluations, introducing an appealing alternative to running costly/lengthy numerical simulations and they mimic the behavior of the high-fidelity simulation model as closely as possible while being computationally cheap to evaluate. 

The paper is structured as follows. In Section \ref{sec:review} we introduce PDEs with random input parameters and stochastic collocation methods. We also describe SGSs and recall the main aspects of the numerical solution via embedded ensemble propagation. In Section \ref{sec:Grouping} we describe the construction of the surrogates for the number of iterations using sparse grid approximations and show how to use them for sample grouping. We also report the results of analytic test cases that illustrate the proposed approach. In Section \ref{sec:numerical-tests} we present the results of numerical tests for anisotropic diffusion problems in three-dimensional spatial domains and multi-dimensional parameter spaces. Here we demonstrate the efficacy of the surrogate-based grouping and its overall better performance with respect to the parameter-based grouping approach. Finally, in Section \ref{sec:conclusion}, we draw conclusions and present future research plans.

\section{Preliminaries}\label{sec:review}
In this section we briefly introduce stochastic PDEs (SPDEs) and specify the mathematical model and its uncertain parameters. Following \cite{Gunzburger_14a} we introduce stochastic collocation methods and sparse grid approximations. Also, based on \cite{Phipps_2015} we recall the principal aspects of embedded ensemble propagation for the solution of groups of parameter dependent deterministic PDEs.

\subsection{PDEs with random input parameters}
Let $D\subset\mbRd$ ($d=1,2,3$) be a bounded domain with boundary $\partial D$ and let $(\omg,\mcF,\mbP)$ be a complete probability space\footnote{Here, $\omg$ is a set of realizations, $\mcF$ is a $\sigma$-algebra of events and $\mbP:\mcF\rightarrow[0,1]$ is a probability measure.}. We consider the following stochastic elliptic boundary value problem. Find $u:\overline D\times\omg$ such that almost surely we have that
\begin{equation}\label{eq:ellipticSPDE}
\left\{\begin{array}{rlll}
\mcL(a)u & = & f & \xb\in D \\[.5mm]
\mcB u   & = & g & \xb\in\partial D,
\end{array}\right.
\end{equation}
where $\mcL$ is an elliptic operator defined on $D$ and parametrized by the uncertain parameter $a(\xb,\omega)$, and $f(\xb)$ is a forcing term with $\xb\in D$ and $\omega\in\omg$. $\mcB$ is a boundary operator and $g(\xb)$ is a boundary data with $\xb\in\partial D$\footnote{For details regarding the functional spaces and the well-posedness of problem \eqref{eq:ellipticSPDE} we refer to \cite{Gunzburger_14a}.}.
We make the following assumptions.
\begin{enumerate}
\item $a(\xb,\omega)$ is bounded from above and below with probability 1. \label{assump1}
\vspace{.05cm}
\item $a(\xb,\omega)$ can be written as $
       a(\xb,\omega) = a(\xb, \yb(\omega))$ in $\overline D\times \omg$,
       where $\yb(\omega) = (y_1(\omega) \ldots y_N(\omega))\in\mbR^N$ is a random 
       vector with uncorrelated components.\label{assump2}
\vspace{.05cm}
\item $a(\xb, \yb)$ is $\sigma$-measurable with respect to $\yb$.\label{assump3}
\end{enumerate}
\smallskip\noindent
A classical example of random parameter that satisfies \ref{assump1}--\ref{assump3} is given by a truncated KL expansion \cite{Loeve_1977,Loeve_1978}, i.e.
\begin{equation}\label{eq:truncatedKL}
a(\xb,\omega) = \mathbb E_a + \sum\limits_{n=1}^N 
\sqrt{\lambda_n} \, b_n(\xb)y_n(\omega).
\end{equation}
The latter corresponds to the approximation of a second order correlated random field with expected value $\mathbb E_a$ and covariance $cov(\xb,\xb')$ with eigenvalues (in decreasing order) $\lambda_n$ and eigenfunctions $b_n(\xb)$.
Note that the random variables $\{y_n(\omega)\}_{n=1}^N$ map the sample space $\omg$ into $\mbR^N$; for $\Gamma_n = y_n(\omg)\subset\mbR$, we define the parameter space as $\Gamma=\prod_{n=1}^N \Gamma_n$. Also, we denote the probability density function of $\yb$ by $\rho(\yb):\Gamma\rightarrow\mbR_+$ with $\rho\in L^\infty(\Gamma)$.

\smallskip\paragraph{A stochastic linear elliptic PDE} 
In this work we consider the following SPDE in $(D\times\Gamma)\subset(\mbR^d\times\mbR^N)$
\begin{equation}\label{eq:elliptic-SPDE}
\left\{\begin{array}{ll}
\mcL(a(\xb,\yb))u = -\nabla\cdot(A(\xb,\yb)\nabla u) = f & \; \xb\in D, \yb\in\Gamma \\[1mm]
\mcB u        =                    u = 0 & \; \xb\in \partial D 
\end{array}\right.
\end{equation}
where $f\in L^2(D)$ is a forcing term and $A(\cdot,\yb):\mbR^N\to\mbR^{d\times d}$ is a diffusivity tensor. As an example, for $d$=3, $A$ may be defined as $A(\xb,\yb)=diag(a(\xb,\yb), \, a_y, \,a_z)$ with $a_y,a_z \in \mbR_+$ and
\begin{equation}\label{eq:diff-coeff}
a(\xb,\yb) = a_{\min} + \widehat a\exp\left\{\sum\limits_{n=1}^N \sqrt{\lambda_n} b_n(\xb) y_n \right\}.
\end{equation}
Here, instead of using the classical KL expansion, to preserve the positive-definiteness of the diffusion tensor required for the well-posedness of problem \eqref{eq:elliptic-SPDE} we consider the expansion of the logarithm of the random field. Again, $\lambda_n$ and $b_n$, $n=1,\ldots N$, are the eigenvalues and eigenfunctions of the covariance function.

\smallskip\paragraph{Quantity of interest}
In the SPDE context the goal of uncertainty quantification is to determine statistical information about an output of interest that depends on the solution. In most of the cases the output of interest is not the solution itself but a functional $G_u(\yb)$, e.g., the spatial average of $u(\cdot,\yb)$: 
$
G_u(\yb)=\frac{1}{|D|} \int_D u(\xb,\yb)\,d\xb
$. 
Then, the statistical information may come in the form of moments of $G_u(\yb)$; as an example, the quantity of interest (QoI) could be the expected value of $G_u(\yb)$ with respect to the probability density function $\rho(\yb)$, i.e.
$
{\rm QoI} = \mathbb E\left[G_u(\yb)\right] = \int_\Gamma G_u(\yb) \rho(\yb)\,d\yb$.
Note that UQ methods aim to find accurate approximations of $G_u$, say $\widehat G_u$, that are then used to cheaply evaluate the QoI.

\subsection{Numerical solution via stochastic collocation methods}\label{sec:lagrange-sparsegrids}
For the finite-dimensional approximation of problem \eqref{eq:elliptic-SPDE} we focus on stochastic collocation (SC) methods; these are non-intrusive stochastic sampling methods based on decoupled deterministic solves.

Given a Galerkin method for spatial discretizations of \eqref{eq:elliptic-SPDE}, we denote by $u_h(\cdot,\yb)$ the semi-discrete approximation of $u(\xb,\yb)$ for all random vectors $\yb\in\Gamma$. The main idea of stochastic collocation methods is to collocate $u_h(\cdot,\yb)$ on a suitable set of samples $\{\yb_m\}_{m=1}^M\subset\Gamma$ to determine $M$ semi-discrete solutions and then use the latter to construct a global or piecewise polynomial to represent the fully SC discrete approximation $u_{hM}^{SC}(\xb,\yb)$, i.e.
\begin{displaymath}
u_{hM}^{SC}(\xb,\yb)=\sum_{m=1}^M c_m(\xb)\psi_m(\yb),
\end{displaymath}
where $\{\psi_m\}_{m=1}^M$ are polynomial basis functions and $c_m(\xb)$ are coefficients that depend on the semi-discrete solutions.
The great advantage of interpolatory approximation is that there is a complete decoupling of spatial and probabilistic discretizations. Also, they are very easy to implement (requiring only codes for deterministic PDEs to be used as black boxes) and embarassingly parallelizable. In this paper we consider only local stochastic collocation methods. In fact, global polynomial approximations perform well only when the solution $u(\xb,\yb)$ is smooth with respect to the random parameters $\{y_n\}_{n=1}^N$ and fail to approximate solutions that have an irregular dependence. Because we are mainly interested in the latter scenario, we resort to local approaches which use locally supported piecewise polynomials to approximate the dependence of the solution on the random parameters.

\smallskip\paragraph{Lagrange interpolation and sparse grids}
We consider a generalized version of sparse grids, introduced by Smolyak in \cite{Smolyak_63}, used in \cite{Back_11,Nobile:2008dr,Nobile:2008kv} that relies on tensor products of one-dimensional approximations. We choose the basis $\{\psi_m\}_{m=1}^M$ to be a piecewise hierarchical polynomial basis \cite{Bungartz_04,Griebel_1998}. These methods achieve higher accuracy by grid refinement in the parameter space, keeping the polynomial degree fixed.

We introduce univariate Lagrange interpolation and then extend it to the multivariate case by tensor products. For simplicity and without loss of generality we consider one-dimensional hat functions defined in $[-1,1]$ as 
\begin{displaymath}
\psi_{l,i}(y) = \psi\left(\frac{y+1-i\,h_l}{h_l}\right), \quad {\rm with} \; \psi(y) = \max\{0,\,1-|y|\},
\end{displaymath}
where, $l=0,1,\ldots$ is the resolution level, $h_l=2^{-l+1}$ is the grid size of level $l$ and $y_{l,i}=i\,h_l-1$, $i=0,1,\ldots 2^l$, are the grid points\footnote{There are several methods for the generation of the set of points within each level. We mention e.g. Gaussian and Clenshaw-Curtis points and we refer to \cite{Gunzburger_14a} for further details.}. Note that this function has local support $(y_{l,i}-h_l,\, y_{l,i}+h_l)$ and it is centered in $y_{l,i}$.

For the space $L^2_\rho(\Gamma)$\footnote{$L^2_\rho(\Gamma)$ is the space of square integrable functions with respect to the probability density function $\rho$.}, we introduce the finite-dimensional subspace of continuous piecewise linear polynomials $Z_l = span\{\psi_{l,i}(y):\; i=0,1,\ldots 2^l \}$, for $l=0,1,\ldots$, and we consider a {\it hierarchical basis}. Let $B_l$, for $l=1,2,\ldots$, be hierarchical index sets defined as $B_l=\{i \in \mbN: \; i=1,3,5,\ldots 2^l-1\}$ and let $W_l$ be the sequence of incremental hierarchical subspaces of $L^2_\rho(\Gamma)$ defined as $W_l = span\{\psi_{l,i}:\; i\in B_l\}$. The hierarchical basis for $Z_l$ is then given by
\begin{displaymath}
\{\psi_{0,0},\psi_{0,1}\} \cup \left\{\bigcup\limits_{l'=1}^l \{\psi_{l',i}(y)\}_{i\in B_{l'}} \right\}.
\end{displaymath}
For each grid level $l$ the interpolant of a function $v\in L^2_\rho(\Gamma)$ in terms of the nodal basis and its {\it incremental interpolation} operator are given by $
\mcI_l(v(y)) = \sum_{i=0}^{2^l} v(y_i) \psi_{l,i}(y)$
and $
\Delta_l(v) = \mcI_l(v) - \mcI_{l-1}(v)$,
respectively. The paper \cite{Gunzburger_14a} shows that the latter can be written in terms of the hierarchical basis functions at level $l$, i.e. 
\begin{displaymath}
\Delta_l(v) = \sum\limits_{i\in B_l} c_{l,i}\, \psi_{l,i}(y), \quad {\rm with} \quad c_{l,i} = v(y_{l,i}) - \mcI_{l-1}(v(y_{l,i})).
\end{displaymath}
We refer to $c_{l,i}$ as {\it surpluses} on level $l$; these quantities play a crucial role in the adaptive generation of the sparse-grid approximation.

\smallskip
For the interpolation of a multivariate function $v(\yb)$ defined on $[-1,\,1]^N$ we extend the one-dimensional hierarchical basis to $N$ dimensions by tensorization. Specifically, we use tensor products to define the basis function associated with the point $\yb_{\lb,\ib}=(y_{l_1,i_1}, \ldots y_{l_N,i_N})$:
\begin{displaymath}
\psi_{\lb,\ib}(\yb) = \prod\limits_{n=1}^N \psi_{l_n,i_n}(y_n),
\end{displaymath}
being $\psi_{l_n,i_n}$ the one-dimensional hierarchical basis function associated with $y_{l_n,i_n} = i_n\,h_{l_n}-1$, for $h_{l_n}=2^{-l_n+1}$; $\lb$ is a multi-index indicating the resolution level along each dimension. Accordingly, we define the $N$-dimensional incremental subspace $W_\lb$ as
\begin{displaymath}
\begin{aligned}
W_\lb & = \bigotimes\limits_{n=1}^N W_{l_n} = span\{\psi_{\lb,\ib}:\; \ib\in B_\lb\}, \quad {\rm where}  \\
B_\lb & = \left\{
\begin{array}{lll}
\ib\in\mbN^N: & \; i_n\in \{1,3,5,\ldots 2^{l_n}-1\} & \;\;  n=1,\ldots N, \; l_n>0 \\[3mm] 
 &              \; i_n\in \{0,1\} &                    \;\;  n=1,\ldots N, \; l_n=0
\end{array}\right\}.
\end{aligned}
\end{displaymath}
Then, we define the sequence of subspaces $Z_l$ as
$
Z_l = \bigoplus\limits_{l'=0}^l \; \bigoplus\limits_{\alpha(\lb')=l'} W_{\lb'},$
where $W_l$ is an incremental subspace and $\alpha$ is a mapping between the multi-index $\lb$ and the level of the sparse-grid approximation, e.g., $\alpha(\lb) = |\lb| = \sum_{n=1}^N l_n$. The latter leads to a sparse polynomial space where the $l$-level hierarchical sparse-grid interpolant of $v(\yb)$ is given by
\begin{equation}\label{eq:hierSGint}
v_l(\yb) = \displaystyle\sum\limits_{l'=0}^l \; \sum\limits_{|\lb'|=l'} \big(\Delta_{l'_1}\otimes \ldots \Delta_{l'_N} \big) v(\yb) \;=\; \displaystyle v_{l-1}(\yb) + \sum\limits_{|\lb'|=l'} \;\sum\limits_{\ib\in B_{\lb'}} c_{\lb',\ib} \psi_{\lb',\ib}(\yb),
\end{equation}
where $c_{\lb',\ib}=v(\yb_{\lb',\ib})-v_{l'-1}(\yb_{\lb',\ib})$ are the $N$-dimensional hierarchical surpluses. The corresponding set of sparse-grid points is then given by $\msH_{\;\lb}(\Gamma) = \{\yb_{\lb,\ib}:\; \ib\in B_\lb\}$. Thus, the sparse grid associated with $v_l$ is given by
$
\msH_l^N(\Gamma) = \bigcup_{l'=0}^l \; \bigcup_{|\lb'|=l'}\msH_{\;\lb'}(\Gamma),
\quad \hbox{with cardinality} \quad \left|\msH_l^N\right|=M_l.
$

\medskip\paragraph{Construction of the fully discrete approximation}
Given a set of points $\{\yb_m\}_{m=1}^M$ and a corresponding set basis functions, we write the fully discrete approximation as 
\begin{equation}\label{eq:uhL}
u_{hL}(\xb,\yb)= \sum\limits_{l=0}^L \; \sum\limits_{|\lb|=l} \; \sum\limits_{\ib\in B_\lb} c_{\lb,\ib}(\xb) \; \psi_{\lb,\ib}(\yb),
\end{equation}
where the coefficients $c_{\lb,\ib}(\xb)$ depend on the finite element solutions corresponding to the sparse-grid points in $\msH_L^N$. Specifically, they are linear combinations of the spatial finite element basis $\{\phi_j(\xb)\}_{j=1}^J$ (being $J$ the number of degrees of freedom), i.e. $c_{\lb,\ib}(\xb) = \sum_{j=1}^J c_{j,\lb,\ib}\,\phi_j(\xb)$. Thus, we can rewrite \eqref{eq:uhL} as
\begin{equation}\label{eq:uhLj}
u_{hL}(\xb,\yb)= \sum\limits_{j=1}^J \left( \sum\limits_{l=0}^L \; \sum\limits_{|\lb|=l} \; \sum\limits_{\ib\in B_\lb} c_{j,\lb,\ib} \; \psi_{\lb,\ib}(\yb)\right) \phi_j(\xb).
\end{equation}
Given the $M_L$ finite element solutions 
$u_h(\xb_j, \yb_{\lb,\ib})$, for $j=1,\ldots J$, $|\lb|\leq L$, and $\ib\in B_\lb$, 
the surpluses $\{c_{j,\lb,\ib}\}$ can be obtained solving the triangular linear system\footnote{The triangular structure of the system is a consequence of the hierarchical nature of the basis, which satisfies $\psi_{\lb,\ib}(\yb_{\lb',\ib'})=0$ if $l'\leq l$.}
\begin{displaymath}
u_{hL}(\xb_j,\yb_{\lb',\ib'}) = \sum\limits_{l=0}^L \; \sum\limits_{|\lb|=l} \; \sum\limits_{\ib\in B_\lb} c_{j,\lb,\ib} 
\; \psi_{\lb,\ib}(\yb_{\lb',\ib'}) = u_h(\xb_j,\yb_{\lb',\ib'}), \quad {\rm for} \; |\lb'|\leq L, \; {\rm and} \; \ib\in B_\lb.
\end{displaymath}

\paragraph{Adaptivity}
Note that using the properties of the hierarchical surpluses \cite{Gunzburger_14a} we can rewrite the approximation \eqref{eq:uhLj} in a hierarchical manner:
\begin{displaymath}
u_{hL}(\xb,\yb) = u_{h(L-1)}(\xb,\yb) + \Delta u_{hL}(\xb,\yb)
\end{displaymath}
where $u_{h,L-1}$ is the sparse-grid approximation in $Z_{L-1}$ and $\Delta u_{h,L}$ is the hierarchical surplus interpolant in the subspace $W_L$ obtained by tensorization. In \cite{Bungartz_04} Bunzgart and Griebel show that for smooth functions the surpluses $c_{j,\lb,\ib}$ of the sparse-grid interpolant $u_{h,L}$ are such that $c_{j,\lb,\ib}\to 0$ as $l\to\infty$. As a consequence, the magnitude of the surpluses can be used as an error indicator for the construction of adaptive sparse-grid interpolants; this technique is particularly powerful with irregular functions, featuring e.g. steep slopes or discontinuities. 

In one dimension the adaptive construction of the sparse grid is straightforward. At each successive interpolation level the surpluses $c_{j,l,i}$, for $j=1,\ldots J$ are evaluated at the points $y_{l,i}$, for $i\in B_l$; if $\max_j|c_{j,l,i}|\geq \tau$, then the grid is refined around $y_{l,i}$ adding the two neighbor points. Here, $\tau$ is a prescribed error tolerance. 

We generalize this strategy to the $N$-dimensional case keeping in mind that each grid point has $2N$ children at each successive level. On each level $l$ we define the new set of indexes $B_\lb^\tau$, for $|\lb|=l$, as
$
B_\lb^\tau = \left\{ \ib\in B_\lb: \; \max_{j=1,\ldots J} |c_{j,\lb,\ib}|\geq\tau \right\}.$
This set only contains the indexes of the surpluses with magnitude larger than $\tau$ for all $j=1,\ldots J$; we refer to this strategy as {\it classic refinement}. This algorithm does not necessarily result in a stable interpolant, i.e. it may fail to converge as $l\to\infty$. Such instabilities may be caused by situations in which $\yb_{\lb,\ib}$ is associated with a large surplus while some of its parents are not included in the point set at the previous level. For this reason, other refinement techniques (based on the classic refinement) have been considered, see \cite{Stoyanov_13} for a summary of alternative adaptive-refinement strategies and their implementation.
\begin{remark}
As already mentioned, the output of interest is usually a functional of the solution; in such cases at each step of the adaptive grid generation we construct an approximation, or surrogate, of $G_u(\yb)$ and base the stopping criterion on the accuracy of the surrogate itself. 

Also, for the purpose of this paper, it must be noted that at each step we can compute multiple surrogates for functionals of interest, i.e., approximations of $I_u(\yb)$, to be used for different tasks; in our case, a functional representing the number of linear solver iterations required for the computation of $u_h$ is computed to perform the grouping.
\end{remark}

\subsection{Numerical solution via ensembles}\label{sec:embedded-propagation}
Sampling-based uncertainty quantification methods such as the stochastic collocation methods described above are attractive since they can be applied to any scientific simulation code with little to no modification of the code.  Furthermore, these methods are trivially parallelizable since each sample can be evaluated independently and therefore in parallel.  However in many cases of interest to large-scale scientific computing, each sample evaluation consumes a large fraction of the available computational resources due to the extremely high fidelity and complexity of the simulations.  Therefore it is often possible to only parallelize a small fraction of the required sample evaluations, with the remaining fraction evaluated sequentially.  Moreover, in many cases a large amount of data and computation is the same in each sample evaluation and in principle could be reused across samples that are being evaluated sequentially, potentially reducing aggregate computational cost.

In this context, an intrusive sample propagation scheme called {\em embedded ensemble propagation} was introduced in~\cite{Phipps_2015} where small groups of samples (called ensembles) are propagated together through the simulation.  Given a user-chosen ensemble size $S$ (typically in the range of 4-32), this approach requires modifying the simulation code to replace each sample-dependent scalar with a length-$S$ array and mapping arithmetic operations on those scalars to the corresponding operation on each component of the array.  In~\cite{Phipps_2015} it was demonstrated that this approach can substantially reduce the cost of evaluating $S$ samples compared to evaluating them sequentially for several reasons:
\begin{itemize}
\item Sample independent quantities (for example spatial meshes and sparse matrix graphs are often sample independent) are automatically reused.  This reduces computation by only computing these quantities once per ensemble, reduces memory usage by only storing them once per ensemble, and reduces memory traffic by only loading/storing them once per ensemble.
\item Random memory accesses of sample-dependent quantities are replaced by contiguous accesses of ensemble arrays.  This amortizes the latency costs associated with these accesses over the ensemble, since consecutive memory locations can usually be accessed with no additional latency cost.  It was demonstrated in \cite{Phipps_2015} that this effect, combined with reuse of the sparse matrix graph can result in 50\% reduction in cost of matrix-vector products associated with sparse iterative linear system solvers on emerging computational architectures, when applied to scalar diffusion problems such as those considered here.
\item Arithmetic on ensemble arrays can be naturally mapped to fine-grained vector parallelism present in most computer architectures today, and this vector parallelism can be more easily extracted by compilers than can typically be extracted from the simulation itself.
\item The number of distributed memory communication steps of sample-dependent information (e.g., within sparse iterative linear system solvers) is reduced by a factor of $S$, with the size of each communication message increased by a factor of $S$.  This both reduces the latency cost associated with these messages by $S$ as well as improves the throughput of each message since larger messages can often be communicated with higher bandwidth.  It was demonstrated in \cite{Phipps_2015} that this can substantially improve scalability to large processor counts when the costs associated with distributed memory communication become significant.
\end{itemize}
Furthermore, it was also shown in \cite{Phipps_2015} that the translation from scalar to ensemble propagation within C++ simulation codes can be facilitated through the use of a template-based generic programming approach~\cite{Pawlowski:2012kc,Pawlowski:2012js} whereby the traditional floating point scalar type is replaced by a template parameter.  This template code can then be instantiated on the original floating point type to recover the original simulation, as well as a new C++ {\em ensemble scalar type} that internally stores the length-$S$ ensemble array to implement the ensemble propagation.  Such a scalar type is provided by the Stokhos~\cite{StokhosURL} package within Trilinos~\cite{TrilinosTOMS,TrilinosSP} and has been integrated with the Kokkos~\cite{Kokkos:2012:SciProg,Kokkos:2014:JPDC} package for portable shared-memory parallel programming as well as the Tpetra package~\cite{Baker:2012:SciProg} for distributed linear algebra.

In~\cite{Phipps_2015} it was shown that the ensemble propagation method was equivalent to solving commuted, Kronecker product systems.  To be precise, consider a finite element discretization of \eqref{eq:elliptic-SPDE}. For every sample $\yb_m$, $m=1,\ldots M$, we write the resulting algebraic system as follows
\begin{equation}\label{eq:discrete-sys}
L_m \Ub_m = \Fb_m, \quad L_m\in \mbR^{J\times J},\; \Ub_m\in\mbR^J,\; \Fb_m\in\mbR^J,
\end{equation}
where $J$ is the number of spatial degrees of freedom\footnote{Note that here we allow the forcing term $f$ to be sample dependent.}.
Consider solving~\eqref{eq:discrete-sys} for $S$ samples $\yb_{m_1},\dots,\yb_{m_S}$:
\begin{equation}\label{eq:discrete-ensemble-sys}
\begin{aligned}
L_{m_1} \Ub_{m_1} &= \Fb_{m_1}, \\
&\vdots \\
L_{m_S} \Ub_{m_S} &= \Fb_{m_S}, \\
\end{aligned}
\end{equation}
which can be written more compactly through Kronecker product notation:
\begin{equation}\label{eq:discrete-kron-sys}
\left(\sum_{i=1}^S e_ie_i^T \otimes L_{m_i}\right)\left(\sum_{i=1}^S e_i\otimes \Ub_{m_i}\right) = \sum_{i=1}^S e_i \otimes \Fb_{m_i}.
\end{equation}
Here $e_i$ is the $i$-th column of the $S\times S$ identity matrix.  Furthermore, a symmetric permutation may be applied to~\eqref{eq:discrete-kron-sys} which results in commuting the order of the terms in each Kronecker product:
\begin{equation}\label{eq:discrete-comm-kron-sys}
\left(\sum_{i=1}^S L_{m_i} \otimes e_ie_i^T \right)\left(\sum_{i=1}^S \Ub_{m_i} \otimes e_i \right) = \sum_{i=1}^S \Fb_{m_i} \otimes e_i.
\end{equation}
Systems~\eqref{eq:discrete-kron-sys} and~\eqref{eq:discrete-comm-kron-sys} are mathematically equivalent, but have different orderings of degrees of freedom.  In~\eqref{eq:discrete-kron-sys}, all spatial degrees of freedom for a given sample $\yb_{m_i}$ are ordered consecutively, whereas in~\eqref{eq:discrete-comm-kron-sys} degrees of freedom for all samples are ordered consecutively for a given spatial degree of freedom.

The embedded ensemble propagation method described in~\cite{Phipps_2015} produces linear systems equivalent to the commuted Kronecker product system~\eqref{eq:discrete-comm-kron-sys} by storing each nonzero entry $(i,j)$ in the ensemble matrix as a length-$S$ array $\{(L_{m_1})_{ij},\dots,(L_{m_S})_{ij}\}$.  Furthermore, to maintain consistency with the Kronecker-product formulation, norms and inner products of ensemble vectors produce scalar results by summing the components for the norm/inner-product across the ensemble.  In terms of sparse iterative solvers such as the conjugate gradient (CG), this has the effect of coupling the systems in~\eqref{eq:discrete-ensemble-sys} together, causing them all to converge at the same rate.

Note that this makes it impossible to determine when each system would have converged when solved independently, which is required for the surrogate-based grouping strategy described below.  To remedy this, we changed the implementation of norms and inner products to not sum contributions across the ensemble, and instead compute an ensemble norm/inner-product.  This breaks the equivalency to a Kronecker-product formulation, but is instead equivalent to solving the systems independently as in~\eqref{eq:discrete-sys} and allows each component system to converge at its own rate.  However since the component systems are stored through the ensemble arrays, the iterative solver must continue until all systems have converged.  Through a custom implementation of the iterative solver convergence test, we are able to determine when each system would have converged when solved independently.  How this is implemented in the software is described in the numerical tests section.

\begin{figure}[t]
\centering
\includegraphics[width=0.6\textwidth]{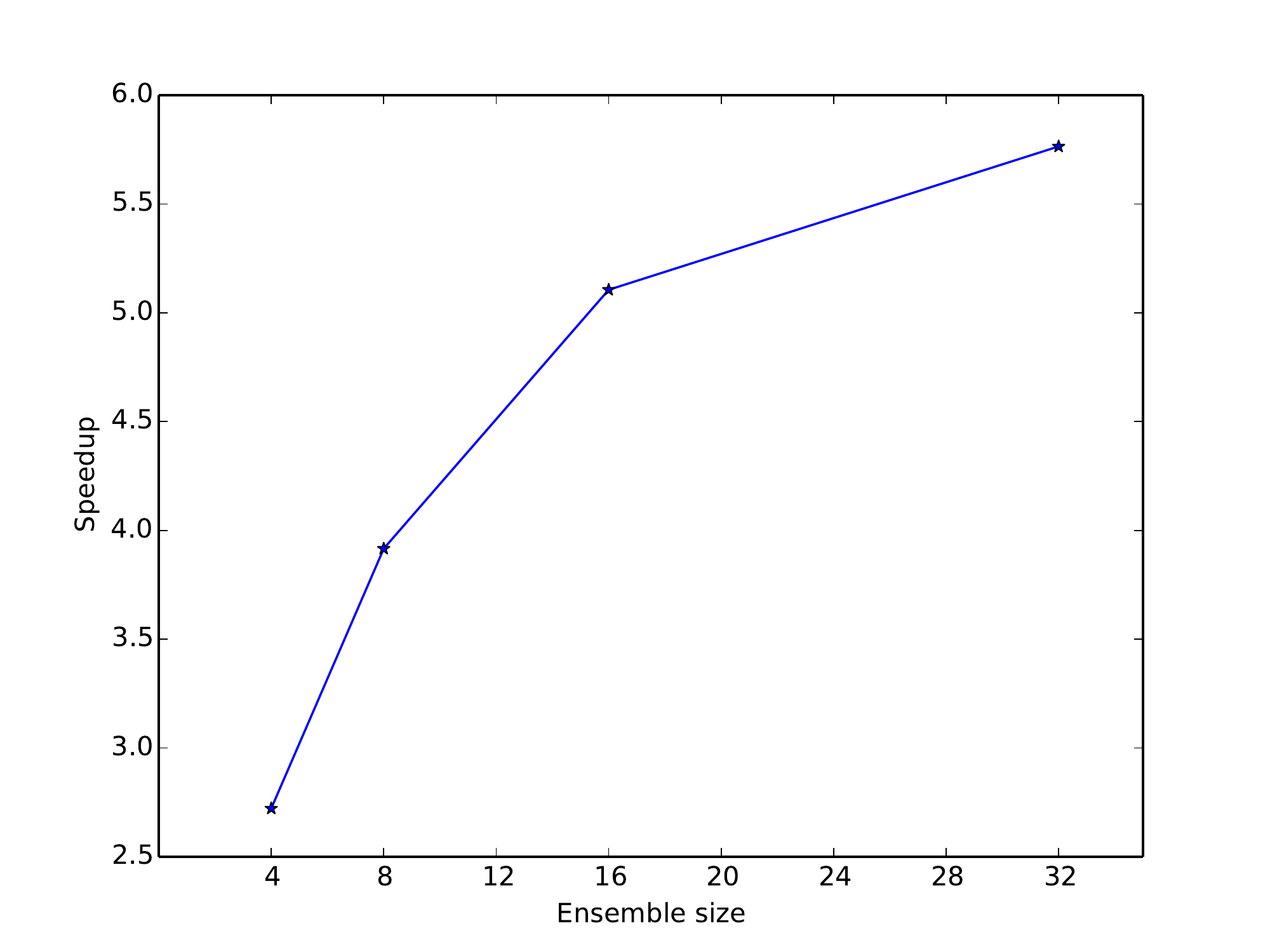}
\caption{Speed-up for the embedded ensemble propagation approach for various ensemble sizes $S$ when applied to a simple isotropic diffusion problem where the number of solver iterations is sample independent, implemented on a single node of the Titan architecture.}
\label{fig:speedup}
\end{figure}

The amount of performance improvement enabled by the embedded ensemble propagation approach is highly problem, problem size, and computer architecture dependent.  For an in-depth examination of performance, see~\cite{Phipps_2015}.  However as motivation for the usefulness of the ensemble propagation approach, as well as to provide a quantitative means of evaluating the impact of the grouping approaches described below, Figure~\ref{fig:speedup} displays the speed-up observed when solving~\eqref{eq:discrete-sys} using the ensemble technique for several choices of ensemble size $S$ relative to solving $S$ systems sequentially.  In these calculations an isotropic diffusion parameter is modeled by the truncated KL expansion $a(\xb,\yb) = a_{\min}+\widehat a\sum_{n=1}^N \sqrt{\lambda_n} b_n(\xb)y_n$, where $\lambda_n$ and $b_n$ are the eigenvalues and eigenfunctions of an exponential covariance, see Section \ref{sec:numerical-tests}, and $y_n\in[-1,1]$.  A spatial mesh of $32^3$ mesh cells was used for the spatial discretization and the resulting linear equations are solved by CG preconditioned with algebraic multigrid (AMG).  The calculations were implemented on a single node of the Titan CPU architecture (16 core AMD Opteron processors using 2 MPI ranks and 8 OpenMP threads per MPI rank).  In this case, due to the isotropy and the fact that we are using a uniform grid, the number of CG iterations is independent of the sample value and therefore the number of CG iterations for each ensemble is independent of the choice of which samples are grouped together in each ensemble.  Essentially, this curve indicates the maximum speed-up possible for the ensemble propagation approach (for the given problem, problem size, and computer architecture) with perfect grouping, obtained when all samples within every ensemble require exactly the same number of preconditioned CG iterations.  Variation in the number of iterations will reduce this speed-up due to increased computational work, which the grouping approaches discussed next attempt to mitigate.

\section{Grouping strategies}\label{sec:Grouping}
In this section we introduce an ensemble grouping approach with the goal of maximizing the performance of the embedded ensemble propagation algorithm introduced in the previous section when the number of linear solver iterations varies dramatically from sample to sample; we describe the grouping algorithm and apply it to analytic test cases for illustration. Also, for the sake of comparison, we recall the parameter-based grouping strategy, a successful technique introduced in \cite{DElia:2016} where the grouping depends on the diffusion parameter characterizing the PDE.

\subsection{Surrogate-based grouping}\label{sec:grouping}
\begin{figure}[t]
\centering
\includegraphics[width=.6\textwidth]{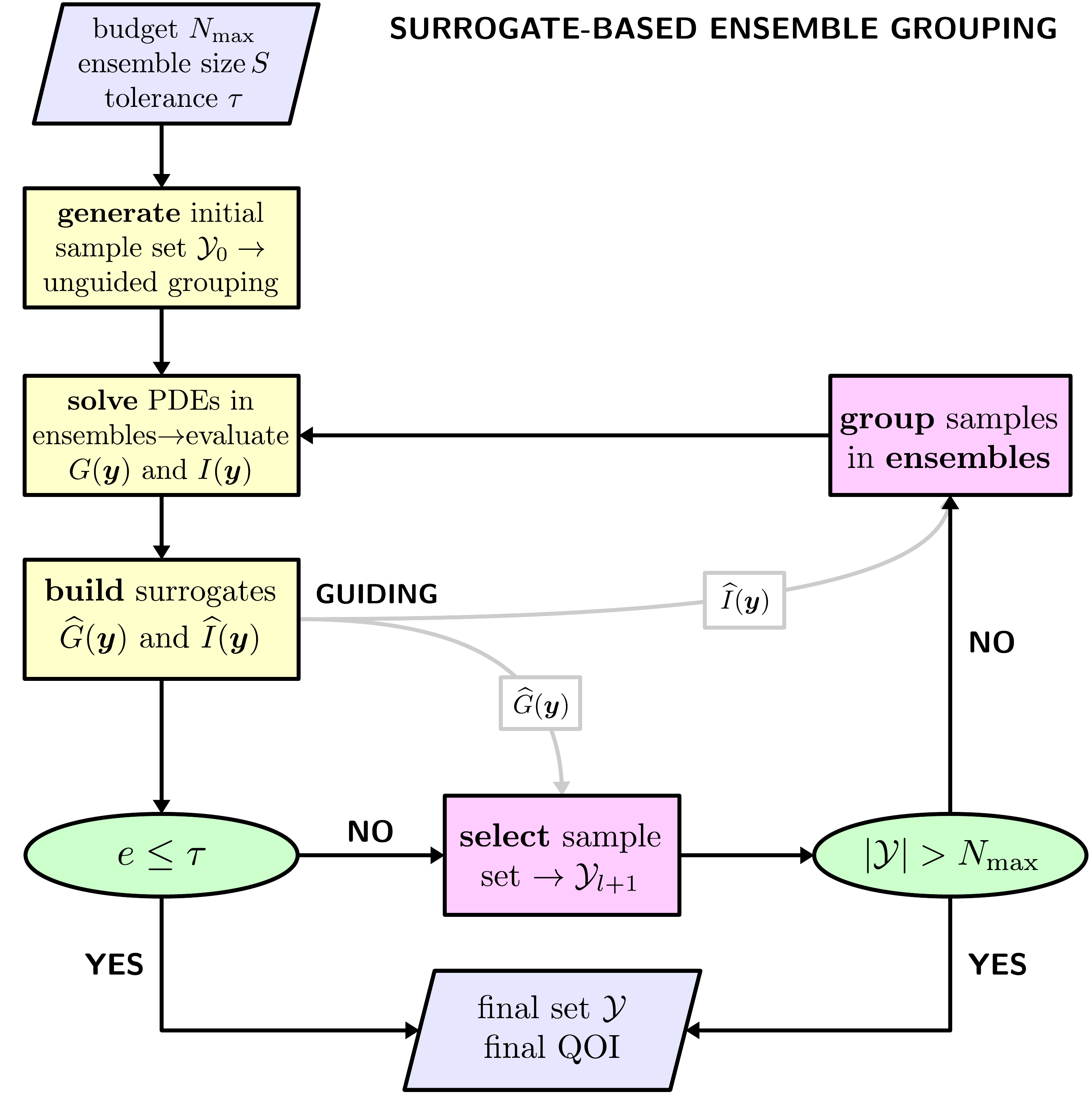}
\caption{Flow chart summarizing the general grid generation and grouping algorithm; here $\mcY=\cup_{l=0}^L \mcY_l$, $N_{\max}$ is the maximum number of samples that one can afford and $\tau$ is a user-defined error tolerance.}
\label{fig:gridGen-grouping-flowchart}
\end{figure}

The key idea of our grouping strategy is to construct a surrogate for the number of iterations so to predict which samples induce a similar convergence behavior and group them together at each step of the adaptive grid generation. In what follows we denote by $G(\widetilde\yb)$ the {\it exact} value of the output of interest at sample $\widetilde\yb$ and by $\widehat G(\widetilde\yb)$ its {\it predicted} value, i.e. $\widehat G(\cdot)$ is a surrogate for the output of interest. Also, we denote by $I(\widetilde\yb)$ the {\it exact} number of iterations associated with sample $\widetilde\yb$ and by $\widehat I(\widetilde\yb)$ the {\it predicted} number of iterations, i.e. $\widehat I(\cdot)$ is a surrogate for the number of linear solver iterations at any point in the sample space. For a general surrogate model we summarize the grid generation and grouping algorithm in Figure \ref{fig:gridGen-grouping-flowchart} and provide more details in the two following paragraphs.
%
\paragraph{The algorithm}
Given a sample budget $N_{\rm max}$, that represents the maximum number of samples that one can afford, an ensemble size $S$, an error tolerance $\tau$, and an error indicator $e$ for the accuracy of the surrogate $\widehat G$, we generate an initial sample set $\mcY_0$ (a sparse grid in our case) and group the samples in ensembles in the order they are generated. Then, we iterate performing the following steps until one of the two stopping criteria (green circles in Figure \ref{fig:gridGen-grouping-flowchart}) is satisfied.
\medskip\begin{itemize}
\item[\sf 1.] {\sf Solve the PDEs in ensembles and evaluate $G(\yb_i)$ and $I(\yb_i)$ for all $\yb_i$ in $\mcY_l$.}

\smallskip\item[\sf 2.] {\sf Build the surrogates $\widehat G$ and $\widehat I$ based on the values of $G$ and $I$ at $\yb_i\in \cup_{l=0}^L \mcY_l$.}

\smallskip{\it If the current surrogate does not satisfy the accuracy requirement, i.e. $e\geq\tau$}

\smallskip\item[\sf 3.] {\sf Use $\widehat G$ to select the new sample set $\mcY_{l+1}$.}

\smallskip{\it If the total number of samples is below the sample budget, i.e. $|\mathcal Y|<N_{\max}$}

\smallskip\item[\sf 4.] {\sf Update $\mathcal Y$, use $\widehat I$ to group the samples in ensembles and go back to {\sf 1}.}
\end{itemize}

\medskip\noindent Note that the number of ensembles to be solved in {\sf 1} is not known a priori but it depends on the sparse grid generator. In fact, the number of samples at every level is determined by the adaptive grid generation algorithm and it depends on the accuracy of the surrogate $\widehat G$. In the next paragraph we describe how to perform {\sf 3} and {\sf 4} using SGS.

\smallskip\paragraph{SGS-based grouping}
As described in Section \ref{sec:lagrange-sparsegrids}, at every level of the adaptive grid generation we can construct a sparse grid approximation of the solution of \eqref{eq:elliptic-SPDE}; alternatively, we can only compute an approximation, or a surrogate, of an output of interest. We express such surrogate in terms of sparse grid basis functions as follows:
\begin{equation}\label{eq:SGSoutput}
\widehat G(\yb)= \sum\limits_{l=0}^L \sum\limits_{|\lb|=l}
                 \sum\limits_{\ib\in B^\tau_\lb} c_{\lb,\ib}\psi_{\lb,\ib}(\yb),
\end{equation} 
where the coefficients $c_{\lb,\ib}$ depend on values of the output of interest at the sparse grid points, and $B^\tau_\lb\!\subset\!B_\lb$ is a subset of the sparse grid index set at level $l$. More specifically, at each level of the adaptive algorithm every point has $2N$ candidate children (or refinement points); in step {\sf 3}, among those candidates, we define the set of new points as those associated with the index set
$B^\tau_{\lb}=\{ \ib\in B_\lb \,:\, c_{\lb,\ib}\geq\tau\}$. 

With the purpose of improving the efficiency of the embedded ensemble propagation, at each level we also build a SGS, $\widehat I$, for the number of linear solver iterations:
\begin{equation}\label{eq:SGSiterations}
\widehat I(\yb)= \sum\limits_{l=0}^L \sum\limits_{|\lb|=l}
                 \sum\limits_{\ib\in B^\tau_\lb} \widetilde c_{\lb,\ib}\psi_{\lb,\ib}(\yb),
\end{equation} 
where $\widetilde c_{\lb,\ib}$ are linear combinations of the number of iterations associated to each point in the current grid. This surrogate is used in step {\sf 4} to group together samples with similar predicted number of iterations. Our strategy consists in ordering the new selected samples according to increasing values of $\widehat I$ and then dividing them in ensembles of size $S$.

Note that we are using a piecewise polynomial surrogate to approximate a function that takes discrete positive values; this coul potentially lead to inaccurate results and compromise the efficiency of the grouping. However, this choice proves to be successful, as shown in the numerical results section.

\smallskip\paragraph{The computational saving}
To assess the computational saving brought by our grouping strategy we consider the quantities
\begin{equation}\label{eq:Rl}
R_l =\displaystyle\frac{S\sum\limits_{k=1}^{K_l} {\bf I}_k}
                     { \sum\limits_{k=1}^{K_l} \sum\limits_{i=1}^S I(\yb_{k,i})}, 
                     \qquad \hbox{or, equivalently} \qquad 
R_l =\displaystyle\frac{S\sum\limits_{k=1}^{K_l} \max\limits_{i=1,\ldots S} I(\yb_{k,i})}
                     { \sum\limits_{k=1}^{K_l} \sum\limits_{i=1}^S I(\yb_{k,i})}
\end{equation}

\begin{equation}\label{eq:R}
R =\displaystyle\frac{S\sum\limits_{k=1}^K {\bf I}_k}
                     { \sum\limits_{k=1}^K \sum\limits_{i=1}^S I(\yb_{k,i})}, 
                     \qquad \hbox{or, equivalently} \qquad 
R =\displaystyle\frac{S\sum\limits_{k=1}^K \max\limits_{i=1,\ldots S} I(\yb_{k,i})}
                     { \sum\limits_{k=1}^K \sum\limits_{i=1}^S I(\yb_{k,i})}
\end{equation}
where ${\bf I}_k$ is the number of iterations required by the $k$th ensemble, $I(\yb_{k,i})$ is the number of iterations required by the $i$th sample in the $k$th ensemble, $K_l$ is the number of ensembles at level $l$ and $K$ is the total number of ensembles. $R_l$ represents the increase in computational work (as indicated by the number of solver iterations) induced by the ensemble propagation at level $l$, whereas $R$ represents the same quantity over all levels.  This increase in work is mitigated by the computational savings induced by the ensemble propagation technique, referred to as speed-up.  The achieved speed-up in practice is then reduced by a factor of $R$.

Note that the equivalence in \eqref{eq:Rl} and \eqref{eq:R} follows from the implementation of the embedded ensemble propagation described in Section \ref{sec:embedded-propagation}.

\smallskip\paragraph{Illustrative tests}
We perform some illustrative tests using quantities of interest represented by analytic functions. More specifically, using continuous and discontinuous functions we test the efficacy of the surrogate-based grouping. For $N=2$, we consider the following quantities
\begin{equation}\label{analytic-QOIs}
\begin{aligned}
G_1(\yb) & = -e^{-(y_1-1)^2} + e^{-0.8(y_1+1)^2}\, e^{-(y_2-1)^2} + e^{-0.8(y_2+1)}, 
             \;\;(y_1,y_2)\in[-2,2]^2\\[3mm]
G_2(\yb) & = \left\{
\begin{array}{ll}
1   &   y_1^2+y_2^2<r_1\\[2mm]
0   &   r_1\leq y_1^2+y_2^2\leq r_2 \\[2mm]
1   &   y_1^2+y_2^2>r_2,
\end{array}\right. \quad(y_1,y_2)\in\;[0,1]^2,\\[3mm]
I(\yb) & = e^{-a_1^2(y_1-u_1)^2- a_2^2(y_2-u_2)^2} +1.
\end{aligned}
\end{equation}
The functions $G_1$ and $G_2$ (only used in this section for testing purposes) play the role of outputs of interest and are used to perform adaptivity, whereas $I$ plays the role of the number of iterations and it is used to test the effectiveness of the surrogate-based grouping, i.e. we build a surrogate for $I$, we use it to order the samples for increasing predicted values of $I$, and we group the samples in ensembles of size $S$.

\smallskip For the generation of the sparse grid and the adaptive refinement based on $\widehat G$ we use TASMANIAN \cite{Stoyanov_13} (toolkit for adaptive stochastic modeling and non-intrusive approximation), a set of libraries for high-dimensional integration and interpolation, and parameter calibration, sponsored by the Oak Ridge National Laboratory. TASMANIAN implements a wide class of one-dimensional rules (and extends them to the multi-dimensional case by tensor products) based on global and local basis functions. In this work the sparse grid is obtained using piecewise linear local basis functions and classic refinement.

It is common practice to apply the adaptive refinement to a grid of level $l>1$; in these experiments we set the initial level to $l=2$. Also, we set $\tau=5\cdot 10^{-4}$ and $N_{\rm max}=1000$ or 2000 and we use the second definition of $R$ in \eqref{eq:R}. Results are reported in Table \ref{tab:R-anl-q1q2-SGS}, for $G_1$ on the left and $G_2$ on the right. Values of $R$ are very close to 1, the optimal value that corresponds to perfect grouping; this is expected due to the ability of SGS to well approximate smooth functions. However, in our application $I$ is not only discontinuous, but it takes values in $\mathbb N_+$; nontheless, results in the next section show the performance improvement enabled by the SGS-based grouping for SPDEs.
\begin{table}[t]
\centering
\begin{tabular}{rll}
\midrule
$S$ & $N_{\rm max}$ & $R$   \\ \midrule
8   & 2000          & 1.428 \\
16  & 2000          & 1.501 \\
20  & 2000          & 1.470 \\
8   & 1000          & 1.064 \\
16  & 1000          & 1.362 \\ 
20  & 1000          & 1.075 \\ \midrule
\end{tabular}
\hspace{.5cm}
\begin{tabular}{rll}
\midrule
$S$ & $N_{\rm max}$ & $R$   \\ \midrule
8   & 2000          & 1.039 \\
16  & 2000          & 1.059 \\
20  & 2000          & 1.067 \\
8   & 1000          & 1.072 \\
16  & 1000          & 1.112 \\
20  & 1000          & 1.114 \\ \midrule
\end{tabular}
\caption{SGS: For $G_1$ (left) and $G_2$ (right), values of $R$ for different ensemble sizes and maximum number of points.}
\label{tab:R-anl-q1q2-SGS}
\end{table}

\subsection{Parameter-based grouping}\label{sec:parameter-grouping}
We recall that we are interested in the solution of anisotropic diffusion problems; common choices of solvers include PGC with algebraic multigrid (AMG) preconditioners. However, AMG methods exhibit poor performance when applied to diffusion problems featuring pronounced anisotropy; this suggests that a measure of the anisotropy is a good indicator of the solver convergence behavior. In \cite{DElia:2016} we proposed as an indicator of slow convergence (high number of iterations) for $\yb_i$ the quantity
\begin{equation}\label{eq:pb-indicator}
H(\widetilde\yb) = \|r(\xb,\widetilde\yb)\|_{L^\infty(D)}
\quad \hbox{where} \quad r(\xb,\widetilde\yb) = 
\frac{\lambda_{\max}(A(\xb,\widetilde\yb))}{\lambda_{\min}(A(\xb,\widetilde\yb))}.
\end{equation}
As in the surrogate-based grouping, this strategy consists in ordering the samples according to increasing values of $H$ and then dividing them in ensembles of size $S$.

The idea of this approach is to identify the intensity of the anisotropy at each point in the spatial domain with the ratio between the maximum and minimum eigenvalues of the diffusion tensor; the maximum value of this quantity over $D$ then provides a measure of the anisotropy associated with the sample $\widetilde\yb$. Note that the computation of this indicator comes at a cost. In fact, prior to the assembling of the ensemble matrix we need to compute for each sample the diffusion tensor and its eigenvalues. 

\smallskip\paragraph{The algorithm}
Given $N_{\rm max}$, $S$, $\tau$, and $e$, we generate an initial sample set $\mcY_0$. Then, we iterate performing the following steps until one of the two stopping criteria is satisfied.
\medskip\begin{itemize}
\item[\sf 1.] {\sf Evaluate $H(\yb_i)$ for all $\yb_i$ in $\mcY_l$ and group the samples.}

\smallskip\item[\sf 2.] {\sf Solve the PDEs in ensembles.}

\smallskip{\it If the current surrogate does not satisfy the accuracy requirement, i.e. $e\geq\tau$}

\smallskip\item[\sf 3.] {\sf Use $\widehat G$ to select the new sample set $\mcY_{l+1}$.}

\smallskip{\it If the total number of samples is below the sample budget, i.e. $|\mathcal Y|<N_{\max}$}

\smallskip\item[\sf 4.] {\sf Update $\mathcal Y$ and go back to {\sf 1}.}
\end{itemize}

\section{Numerical tests}\label{sec:numerical-tests}
In this section we present the results of numerical tests performed on a spatial domain of dimension $d=3$ and a sample space of dimension $N=4$.
For the solution of the anisotropic diffusion problem \eqref{eq:elliptic-SPDE} we let $D=[0,1]^3$ and $\Gamma=[-1,1]^N$. We consider the following exponential covariance function
\begin{displaymath}
cov(\xb,\xb') = \sigma_0 \exp\left\{-\displaystyle\frac{\|\xb-\xb'\|_1}{ \delta}\right\},
\end{displaymath}
where $\delta$ is the characteristic distance of the spatial domain, i.e. the distance for which points in the spatial domain are significantly correlated. In all our simulations we set $\delta=1/4$ and $\sigma_0=\sqrt{300}$.  We discretize~\eqref{eq:elliptic-SPDE} using trilinear finite elements and $32^3$ mesh cells.  The Kokkos~\cite{Kokkos:2012:SciProg,Kokkos:2014:JPDC} and Tpetra~\cite{Baker:2012:SciProg} packages within Trilinos~\cite{TrilinosTOMS,TrilinosSP} are used to assemble and solve the linear systems for each sample value using hybrid shared-distributed memory parallelism via OpenMP and MPI.  The equations are solved via CG implemented by the Belos package~\cite{Bavier:2012:SciProg} with a linear solver tolerance of $10^{-7}$.  CG is preconditioned via smoothed-aggregation AMG as provided by the MueLu package~\cite{muelu_user}.  A 2nd-order Chebyshev smoother is used at each level of the AMG hierarchy and a sparse-direct solve for the coarsest grid.  The linear system assembly, CG solve, and AMG preconditioner are templated on the scalar type for the template-based generic programming approach to implement the embedded ensemble propagation as described in Section~\ref{sec:embedded-propagation}, allowing the code to be instantiated on \texttt{double} for single sample evaluation and the ensemble scalar type provided by Stokhos~\cite{StokhosURL} for ensembles.  As before, the calculations were implemented on a single node of the Titan CPU architecture (16 core AMD Opteron processors using 2 MPI ranks and 8 OpenMP threads per MPI rank).

For the adaptive grid generation we use TASMANIAN and we consider a classic refinement; we set the initial sparse grid level to $l=1$, $\tau=10^{-3}$ and $N_{\rm max}=2000$.  We choose the $\ell^2$-norm of the vector of the values of the discrete solution at the degrees of freedom as the output of interest, i.e. $G(\yb)=\|{\bf u}(\yb)\|_{\ell^2}^2$, where ${\bf u}(\yb)$ is the discrete solution in correspondence of the sample $\yb$.

The surrogate-based grouping strategy described above requires access to the number of CG iterations for each sample within an ensemble in order to build up the iterations surrogate $\widehat{I}$.  As described in Section~\ref{sec:embedded-propagation} we modified the ensemble propagation implementation to compute inner products and norms of ensemble vectors as ensembles (instead of scalars).  Thus the CG residual norm computed during the CG iteration becomes an ensemble value.  The CG iteration must continue until each ensemble residual norm satisfies the supplied tolerance, and thus all samples within the ensemble must take the same number of iterations.  However we modified the convergence decision implementation in Belos (through partial specialization of the Belos convergence test abstraction on the ensemble scalar type) to keep track of when each sample within the ensemble would have converged, based on its component of the residual norm.  These values are then reported to TASMANIAN to build the iterations surrogate.

Note that because of the high variation in CG iterations from sample to sample present in the following tests, and the requirement that the CG iteration continue until each component of the ensemble satisfies the convergence criteria, numerical underflow in the $A$-conjugate norm calculation may occur for some samples within an ensemble.  When this occurs, the corresponding components of the norm appear numerically as zero, resulting in invalid floating point values (i.e., \texttt{NaN}) in the next search direction.  While the grouping strategies generally mitigate this, it may occur with a poor grouping of samples.   To alleviate this, we also modified the CG iteration logic (again via partial specialization of the Belos iteration abstraction) to replace the update with zero for each ensemble component when the $A$-conjugate norm is zero.  For these samples, the CG algorithm continues until the remaining samples have converged, but the approximate solution no longer changes.

\begin{figure}[t]
\centering
\includegraphics[width=\textwidth]{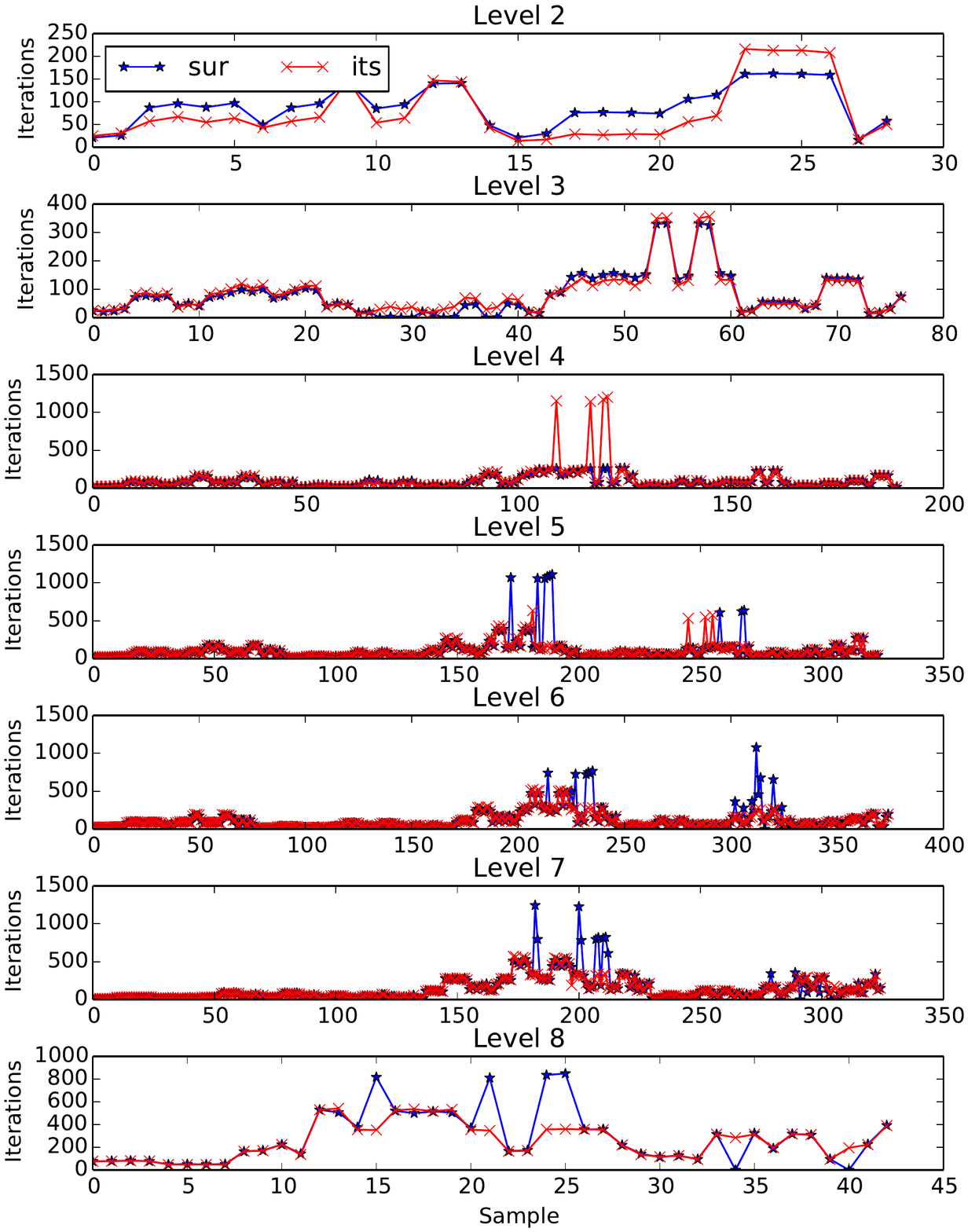}
\caption{Actual (``its'') and surrogate-predicted (``sur'') linear solver iterations for each sample at each adaptive level for Test 1.}
\label{fig:cont-surr}
\end{figure}

\paragraph{Test 1}\label{test1}
For $S=4,\,8,\,16,\,32$ we report the results of our tests in Table \ref{tab:Rvalues_cont}. The adaptive algorithm generates a sparse grids of size $|\mathcal Y|=1372$ after achieving the prescribed error tolerance $\tau$ with eight levels of refinement. The strategies ``sur'', ``par'' and ``nat'' correspond to the SGS-based grouping, the parameter-based grouping and the one based on the order in which the samples are generated by TASMANIAN.  We also include the best hypothetical grouping based on the actual iterations for each sample in the rows labeled ``its''.  For each method and ensemble size, Table \ref{tab:Rvalues_cont} displays the calculated $R_l$ for each level $l$ of the adaptive grid generation (see~\eqref{eq:Rl}), the final $R$ for the entire sample propagation (see~\eqref{eq:R}), and the total measured speedup for the ensemble linear system solves defined to be the time for all linear solves computed sequentially divided by the time for all ensemble solves (for the ``its'' method, a speedup is not computed since it is a hypothetical grouping constructed after all solves have been completed).  To validate the measured speed-ups, we also computed the predicted speed-up determined by the iteration independent speed-up given by Figure~\ref{fig:speedup} divided by $R$.
\begin{table}[t]
\centering
\begin{tabular}{ccccccccccccc}
\midrule
    &     &       &       &       &       &       &       &       &       &      &  Speed- &  Pred. \\
I   & $S$ & $R_1$ & $R_2$ & $R_3$ & $R_4$ & $R_5$ & $R_6$ & $R_7$ & $R_8$ & $R$  & up   & Speed-up  \\ \midrule
its & 4   & 1.68  & 1.43  & 1.23  & 1.05  & 1.01  & 1.04  & 1.06  & 1.10  & 1.06 & --   & 2.56      \\ 
sur & 4   & 2.04  & 1.44  & 1.27  & 1.32  & 1.13  & 1.09  & 1.10  & 1.14  & 1.15 & 2.35 & 2.37      \\ 
par & 4   & 2.04  & 1.71  & 1.52  & 1.30  & 1.34  & 1.20  & 1.12  & 1.12  & 1.24 & 1.81 & 2.20      \\ 
nat & 4   & 2.04  & 1.66  & 1.44  & 1.23  & 1.34  & 1.28  & 1.25  & 1.24  & 1.29 & 2.15 & 2.11      \\ 
\\
its & 8   & 2.83  & 1.60  & 1.27  & 1.08  & 1.10  & 1.08  & 1.10  & 1.44  & 1.14 &  --  & 3.44      \\ 
sur & 8   & 2.83  & 1.67  & 1.33  & 1.14  & 1.13  & 1.11  & 1.12  & 1.44  & 1.17 & 3.35 & 3.35      \\ 
par & 8   & 2.83  & 2.15  & 1.71  & 1.39  & 1.49  & 1.27  & 1.18  & 1.47  & 1.35 & 2.84 & 2.89      \\ 
nat & 8   & 2.83  & 2.29  & 1.90  & 1.48  & 1.49  & 1.51  & 1.41  & 1.64  & 1.52 & 2.61 & 2.58      \\ 
\\
its & 16  & 3.11  & 1.94  & 1.60  & 1.12  & 1.28  & 1.25  & 1.25  & 1.56  & 1.30 &  --  & 3.94      \\ 
sur & 16  & 3.11  & 1.94  & 1.69  & 1.19  & 1.30  & 1.29  & 1.28  & 1.56  & 1.33 & 3.70 & 3.84      \\ 
par & 16  & 3.11  & 2.59  & 1.83  & 1.46  & 1.65  & 1.41  & 1.30  & 1.57  & 1.49 & 3.33 & 3.43      \\ 
nat & 16  & 3.11  & 2.59  & 2.12  & 1.87  & 1.80  & 1.83  & 1.67  & 2.16  & 1.84 & 2.73 & 2.78      \\ 
\\
its & 32  & 6.22  & 3.07  & 2.62  & 1.25  & 1.67  & 1.32  & 1.61  & 2.63  & 1.66 &  --  & 3.46      \\ 
sur & 32  & 6.22  & 3.07  & 2.75  & 1.30  & 1.70  & 1.36  & 1.64  & 2.64  & 1.70 & 3.47 & 3.39      \\ 
par & 32  & 6.22  & 3.07  & 2.76  & 1.59  & 2.11  & 1.57  & 1.65  & 2.64  & 1.87 & 3.05 & 3.08      \\ 
nat & 32  & 6.22  & 3.07  & 2.99  & 2.37  & 2.24  & 2.16  & 2.07  & 2.74  & 2.28 & 2.06 & 2.53      \\
\midrule
\end{tabular}
\caption{Computational results for Test 1, displaying $R_l$ for each level $l$, the final $R$, measured ensemble linear solver speed-up and predicted speed-up based on $R$ and Figure~\ref{fig:speedup}, for the iterations-based (``its''), surrogate-based (``sur''), parameter-based (``par''), and natural grouping (``nat'') methods.}
\label{tab:Rvalues_cont}
\end{table}
Because of the large variation in number of CG iterations from sample to sample, we observe large values of $R$ for the natural ordering based on the order in which samples are generated, particularly for larger ensemble sizes; this reduces the performance of the ensemble propagation method.\footnote{Note that at each level, the number of samples is not usually evenly divisible by the ensemble size.  To use a uniform ensemble size for all ensembles, samples are added by replicating the last sample in the last ensemble.  This results in larger $R$ values when the number of samples is small and the ensemble size is large, as can be seen in the results for $R_1$.} However the parameter and surrogate-based orderings reduce $R$ and therefore lead to larger speed-ups.  The surrogate and iterations-based orderings lead to similar $R$ values, demonstrating that the surrogate approach is predicting the number of solver iterations well.  This is particularly evident for higher refinement levels where a more accurate surrogate for the number of iterations has been constructed.  In Figure~\ref{fig:cont-surr} the number of solver iterations for each sample at each level, as well as the iterations predicted by the surrogate are displayed, demonstrating that the surrogate generally predicts the number of iterations well, and its accuracy generally improves as the stochastic grid is refined.

\begin{figure}[t]
\centering
\includegraphics[width=\textwidth]{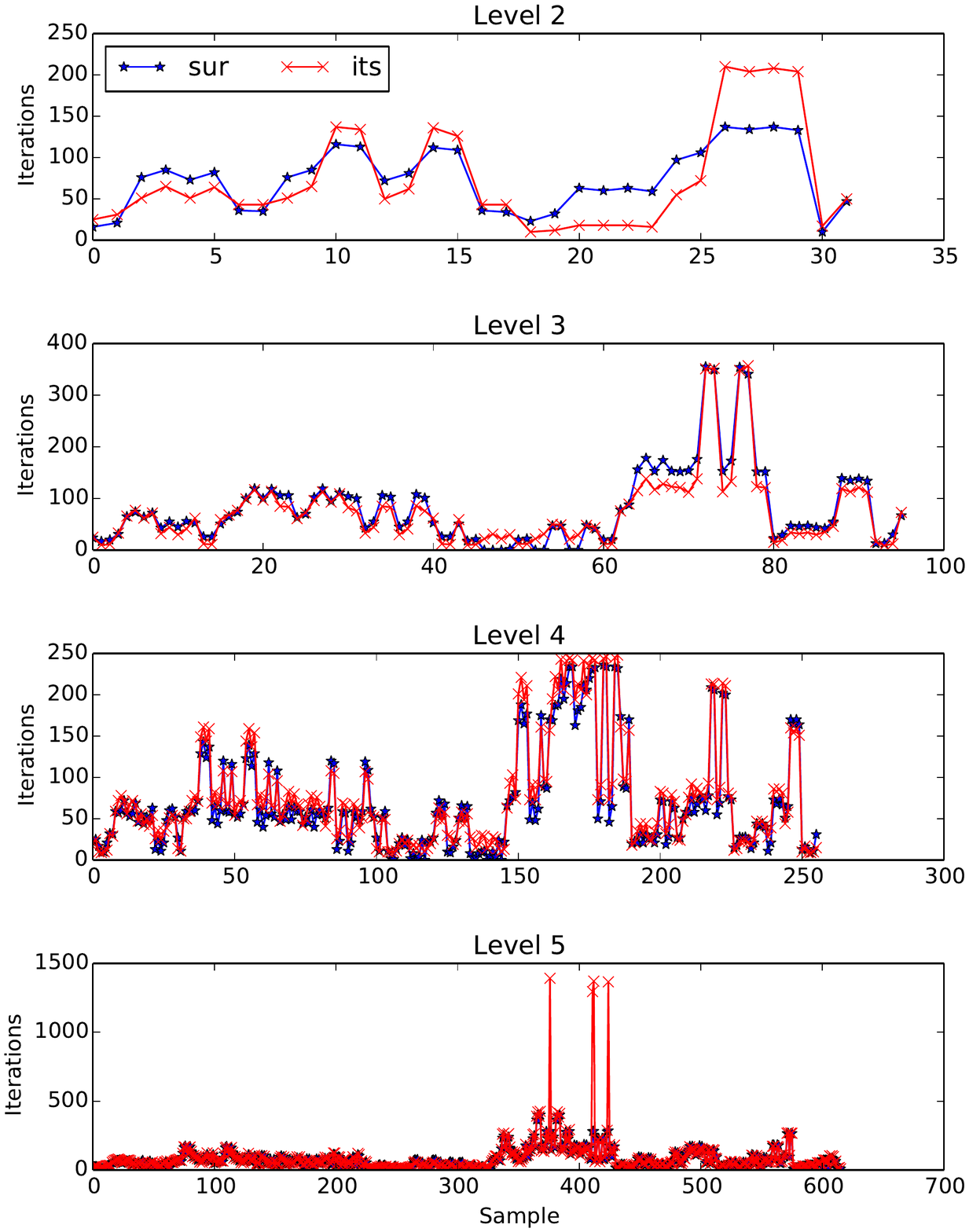}
\caption{Actual (``its'') and surrogate-predicted (``sur'') linear solver iterations for each sample at each adaptive level for Test 2.}
\label{fig:disc-surr}
\end{figure}

\begin{table}[t]
\centering
\begin{tabular}{ccccccccccccc}
\midrule
    &     &       &       &       &       &       &      &  Speed- & Pred.  \\
I   & $S$ & $R_1$ & $R_2$ & $R_3$ & $R_4$ & $R_5$ & $R$  & up   & Speed-up  \\ \midrule
its & 4   & 1.77  & 1.06  & 1.20  & 1.06  & 1.06  & 1.08 & --   & 2.51      \\ 
sur & 4   & 2.08  & 1.13  & 1.24  & 1.14  & 1.25  & 1.22 & 2.09 & 2.23      \\ 
par & 4   & 2.08  & 1.44  & 1.62  & 1.36  & 1.37  & 1.41 & 1.93 & 1.94      \\ 
nat & 4   & 2.08  & 1.51  & 1.32  & 1.34  & 1.35  & 1.36 & 2.00 & 2.01      \\ 
\\
its & 8   & 2.91 & 1.23   & 1.56  & 1.16  & 1.14  & 1.22 &  --  & 3.22      \\ 
sur & 8   & 2.91 & 1.29   & 1.64  & 1.27  & 1.21  & 1.30 & 2.80 & 3.02      \\ 
par & 8   & 2.91 & 1.74   & 2.01  & 1.49  & 1.79  & 1.74 & 2.29 & 2.25      \\ 
nat & 8   & 2.91 & 1.56   & 1.80  & 1.55  & 1.55  & 1.59 & 2.41 & 2.46      \\ 
\\
its & 16  & 3.33 & 1.79   & 1.64  & 1.22  & 1.17  & 1.29 &  --  & 3.97      \\ 
sur & 16  & 3.33 & 1.79   & 1.69  & 1.33  & 1.24  & 1.36 & 3.22 & 3.74      \\ 
par & 16  & 3.33 & 2.38   & 2.37  & 1.60  & 1.66  & 1.77 & 2.87 & 2.88      \\ 
nat & 16  & 3.33 & 2.38   & 2.10  & 1.99  & 1.81  & 1.93 & 2.60 & 2.65      \\ 
\\
its & 32  & 6.65 & 2.88   & 2.28  & 1.38  & 1.28  & 1.54 &  --  & 3.74      \\ 
sur & 32  & 6.65 & 2.88   & 2.34  & 1.46  & 1.37  & 1.62 & 3.04 & 3.55      \\ 
par & 32  & 6.65 & 2.88   & 2.53  & 1.75  & 1.77  & 1.94 & 2.87 & 2.96      \\ 
nat & 32  & 6.65 & 2.88   & 2.87  & 2.56  & 2.16  & 2.43 & 2.39 & 2.38      \\ \midrule
\end{tabular}
\caption{Computational results for Test 2, displaying $R_l$ for each level $l$, the final $R$, measured ensemble linear solver speed-up and predicted speed-up based on $R$ and Figure~\ref{fig:speedup}, for the iterations-based (``its''), surrogate-based (``sur''), parameter-based (``par''), and natural grouping (``nat'') methods.}
\label{tab:Rvalues_disc}
\end{table}
\paragraph{Test 2}
Next we consider a diffusion parameter with a discontinuous behavior with respect to the uncertain variable $\yb$. Specifically, we define $\widehat a$ as follows
\begin{equation}
\widehat a(\yb) = \left\{ 
\begin{array}{ll}
1    & r(\yb)< \frac{d}{4} \\[2mm]
100  & \frac{d}{4}\leq r(\yb)< \frac{d}{2} \\[2mm]
10   & r(\yb) \geq \frac{d}{2},
\end{array}\right.
\end{equation}
with $d=\sqrt{3}$ and $r(\yb)= \sqrt{\sum_i y_i^2}$.  Due to the discontinuous nature of the problem, the adaptive algorithm is unable to achieve the error tolerance $\tau$ within the maximum number of points, and stops after reaching a size of $|\mathcal Y|=1009$ and five levels of refinement. We report the results in Table \ref{tab:Rvalues_disc}; generally, results similar to the continuous case above are observed.  However we do see larger differences between the $R$ values at higher levels between the iteration and surrogate-based groupings.  As before, the solver iterations at each level as well as the iterations predicted by the surrogate are displayed in Figure~\ref{fig:disc-surr}.  Again, the surrogate predicts the number of iterations for most samples reasonably well, even for this more difficult discontinuous case.

\section{Conclusion}\label{sec:conclusion}
The embedded ensemble propagation approach introduced in~\cite{Phipps_2015} has been demonstrated to be a powerful means of reducing the computational cost of sampling-based uncertainty quantification methods, particularly on emerging computational architectures.  A substantial challenge with this method however is ensemble-divergence, whereby different samples within an ensemble choose different code paths.  This can reduce the effectiveness of the method and increase computational cost.  Therefore grouping samples together to minimize this divergence is paramount in making the method effective for challenging computational simulations.  

In this work, a new grouping approach based on a surrogate for computational cost built up during the uncertainty propagation was developed and applied to model diffusion problems where computational cost is driven by the number of (preconditioned) linear solver iterations.  The approach was developed within the context of locally adaptive stochastic collocation methods, where an iterations surrogate generated from previous levels of the adaptive grid generation is used to predict iterations for subsequent samples, and group them based on similar numbers of iterations.  While the approach was developed within the context of stochastic collocation methods, we believe the idea is general and could be easily applied to any adaptive uncertainty quantification algorithm.  In principle it could even be applied to non-adaptive algorithms by pre-selecting a set of samples, evaluating those samples, and generating an appropriate iterations surrogate from those results.  The method was applied to two highly anisotropic diffusion problems with a wide variation in solver iterations from sample to sample, one continuous with respect to the uncertain parameters, and one discontinuous, and the method was demonstrated to significantly improve grouping and increase performance of the ensemble propagation method.  It extends the parametric-based grouping approach developed in~\cite{DElia:2016} to more general problems without requiring detailed knowledge of how the uncertain parameters affect the simulation's cost, and is also less intrusive to the simulation code.

The idea developed here could be further improved by allowing for variation in the ensemble size within each ensemble step.  Given a prediction of how each ensemble size affects performance (e.g., from Figure~\ref{fig:speedup}) and a surrogate for computational cost as developed here, ensembles of varying sizes could be selected to maximize performance through a constrained combinatorial optimization.  Furthermore, the adaptive uncertainty quantification method could be modified to select new points not only based on the PDE quantity-of-interest, but also choose points that minimize divergence of computational cost/iterations.  These ideas will be pursed in future works.

\section{Acknowledgements}
This material is based upon work supported by the U.S. Department of Energy, Office of Science, and Office of Advanced Scientific Computing Research (ASCR),  as well as the National Nuclear Security Administration, Advanced Technology Development and Mitigation program.  This research used resources of the Oak Ridge Leadership Computing Facility, which is a DOE Office of Science User Facility.

The authors would like to thank Dr. Miro Stoyanov for useful conversations and for providing great support with TASMANIAN. 

\bibliographystyle{plain}
\bibliography{paper}

\end{document}